\documentclass[10pt,a4paper,twocolumn,english,prl,showpacs,superscriptaddress]{revtex4-1}

\usepackage[T1]{fontenc}
\usepackage[latin9]{inputenc}
\usepackage{fancyhdr}
\usepackage{color}
\usepackage{babel}
\usepackage{amsmath}
\usepackage{amsthm}
\usepackage{amssymb}
\usepackage{graphicx}
\usepackage[unicode=true,pdfusetitle,
 bookmarks=true,bookmarksnumbered=false,bookmarksopen=false,
 breaklinks=false,pdfborder={0 0 1},backref=false,colorlinks=false]
 {hyperref}

\makeatletter

\pdfpageheight\paperheight
\pdfpagewidth\paperwidth

\IfFileExists{candleshape.def}{%
 \input{candleshape.def}}{}
\IfFileExists{dropshape.def}{%
 \input{dropshape.def}}{}
\IfFileExists{TeXshape.def}{%
 \input{TeXshape.def}}{}
\IfFileExists{triangleshapes.def}{%
 \input{triangleshapes.def}}{}

\hypersetup{
    colorlinks=true,
    linkcolor=red,
    citecolor=blue,
    filecolor=blue,      
    urlcolor=blue,
}
\makeatother

\begin{document}
\title{\noindent \textcolor{black}{Two-level quantum Otto heat engine operating
with unit efficiency far from the quasi-static regime under a squeezed
reservoir }}
\author{Rogério J. de Assis}
\email{rjdeassis@gmail.com}
\address{Instituto de Física, Universidade Federal de Goiás, 74.001-970, Goiânia
- Go, Brazil}
\author{José S. Sales}
\address{Campus Central, Universidade Estadual de Goiás, 75132-903, Anápolis,
Goiás, Brazil}
\author{Udson C. Mendes}
\address{Instituto de Física, Universidade Federal de Goiás, 74.001-970, Goiânia
- Go, Brazil}
\author{Norton G. de Almeida}
\address{Instituto de Física, Universidade Federal de Goiás, 74.001-970, Goiânia
- Go, Brazil}
\begin{abstract}
\textcolor{black}{Recent theoretical and experimental studies in quantum
heat engines show that, in the quasi-static regime, it is possible
to have higher efficiency than the limit imposed by Carnot, provided
that engineered reservoirs are used. The quasi-static regime, however,
is a strong limitation to the operation of heat engines, since infinitely
long time is required to complete a cycle. In this paper we propose
a two-level model as the working substance to perform a quantum Otto
heat engine surrounded by a cold thermal reservoir and a squeezed
hot thermal reservoir. Taking advantage of this model we show a striking
achievement, that is to attain unity efficiency even at non null power. }
\end{abstract}
\maketitle

\section*{Introduction}

Classical heat engines convert thermal resources into work, which
is maximized for reversible operations where the entropy production
vanishes. In the quantum realm, both the engine and the reservoirs
can be composed of finite-dimensional systems \cite{Kieu2004,Quan2007,Wang2009,Linden2010,Scully2011,Wang2012,Rahav2012,Gelbwaser-Klimovsky2013,Uzdin2015,Kosloff2017,Huang2017,Zhao2017,Brandner2017,Niedenzu2018,Dorfman2018,Camati2019,Turkpence2019,Peterson2019,Bera2019,Cakmak2020},
thus departing from classical scenarios. Differently from the classical
case, quantum engines, as well as their environments, can be prepared
in physical states without classical analogues, allowing the efficiency
to be strongly enhanced \cite{RoBnagel2014,Manzano2016,Klaers2017,Assis2019}.
For cyclic heat engines to be useful in the real world, it is necessary
that the cycles be performed in finite time or non-null power, which
introduces losses due to irreversibility. Attempts to build thermal
engines operating with finite-time cycles without losing efficiency
have been made. Some of these attempts consist in exploiting quantum
resources such as coherence \cite{Scully2003}, while others employ
techniques based on the shortcut to adiabaticity \cite{Guery-Odelin2019},
which consists of changing the dynamics of the system to minimize
the irreversible losses of the fast dynamics \cite{Beau2016}. Yet,
other approaches propose nonequilibrium reservoirs as the environment
for heat engines to operate \cite{Xiao2018,Assis2019,Wang2019,Singh2020}.
So far, all those proposals are mainly concerned with investigating
the trade-off between efficiency and power either by trying to eliminate
the irreversible losses during the finite time operation or trying
to get efficiency at maximum power in the quasi-static cycle, which
can be higher than Carnot efficiency \cite{RoBnagel2014}. Exception
to these approaches is the recent theoretical and experimental work
in Ref.~\onlinecite{Assis2019}, where the authors explored reservoirs
with effective negative temperatures and showed that, for a given
set of parameters, the faster the processes are performed, the greater
the efficiency of the engine. 

I\textcolor{black}{n this paper we study the efficiency of a quantum
Otto heat engine (QOHE) with a two-level system (TLS) as a working
substance operating under a cold thermal reservoir and a squeezed
hot thermal reservoir. We then explore how advantageous is to extract
work both at null and at non-null power regime. By comparing our model
with a similar one in which a harmonic oscillator (HO) is used as
the working substance in the QOHE \cite{RoBnagel2014}, we show that
our model presents enhancement in the efficiency at maximum power.
Here we also consider a set of QOHE parameters for a possible experimental
implementation in the field of nuclear magnetic resonance (NMR).}

\section*{The QOHE}

\textcolor{black}{Our model consists of a TLS as the working substance
undergoing unitary operations and interacting with a cold thermal
reservoir and with a squeezed hot thermal reservoir to perform a QOHE.
A QOHE }consists of two isochoric branches, one with the working substance
coupled to the cold thermal reservoir and the other coupled to the
hot thermal reservoir, and two isentropic branches, in which the working
substance is disconnected from the thermal reservoirs and evolves
unitarily. Inspired by recent studies on QOHEs in NMR, Refs.~\onlinecite{Peterson2019,Assis2019}, we consider
here the following four-stroke QOHE:

\emph{(i) Cooling stroke.} Initially, the TLS is weakly coupled to
the cold thermal reservoir up to thermalization. The thermalized TLS
is then described by the Gibbs state $\rho_{1}^{G}=\text{e}^{-\beta_{c}H_{c}}/\text{Tr}\left(\text{e}^{-\beta_{c}H_{c}}\right)$,
where $H_{c}$ is the TLS Hamiltonian and $\beta_{c}=1/k_{B}T_{c}$,
where $k_{B}$ is the Boltzmann constant and $T_{c}$ is the reservoir
temperature. The Hamiltonian has the form $H_{c}=\frac{1}{2}\hbar\omega_{c}\sigma_{x}$,
with $\hbar$, $\omega_{c}$, and $\sigma_{x}$ being the reduced
Planck constant, the angular frequency, and the x Pauli matrix, respectively.
The TLS Hamiltonian remains unchanged during the thermalization process.

\emph{(ii) Expansion stroke.} In this stage, from time $t=0$ to $t=\tau$,
the TLS evolves unitarily from the state $\rho_{1}^{G}$ to $\rho_{2}=U\rho_{1}^{G}U^{\dagger}$,
where $U$ is the unitary operator. The unitary evolution is characterized
by the driving of the TLS Hamiltonian from $H_{c}=\frac{1}{2}\hbar\omega_{c}\sigma_{x}$
to $H_{h}=\frac{1}{2}\hbar\omega_{h}\sigma_{y}$, with $\omega_{h}$
being an angular frequency greater then $\omega_{c}$ (energy gap
expansion) and $\sigma_{y}$ the y Pauli matrix. Here, we are not
concerned with the specific form of the unitary operator $U$.

\emph{(iii) Heating stroke.} Here, the TLS is weakly coupled to the
squeezed (non-equilibrium) hot thermal reservoir until reaching the
stead state $\rho_{3}^{S}=S\rho_{3}^{G}S^{\dagger}$. The TLS would
reach the Gibbs state $\rho_{3}^{G}=\text{e}^{-\beta_{h}H_{h}}/\text{Tr}\left(\text{e}^{-\beta_{h}H_{h}}\right)$
when thermalizing with the hot thermal reservoir without squeezing.
However, reservoir squeezing changes the asymptotic state of the TLS
according to operator $S=\left(\mu\left|-_{y}\right\rangle \left\langle +_{y}\right|+\nu\left|+_{y}\right\rangle \left\langle -_{y}\right|\right)/\sqrt{\mu^{2}+\nu^{2}}$,
where $\mu=\cosh r$ and $\nu=\sinh r$. The state $\left|\pm_{y}\right\rangle $
is the eigenstate of $H_{h}$ with eigenenergy $\pm\frac{1}{2}\hbar\omega_{h}$,
and $r$ is the \emph{squeezing parameter}. The form of $\rho_{3}^{S}$
described here can be easily verified by noting its correspondence
with Eq. (10) from Ref.~\onlinecite{Srikanth2008}. As in the cooling stroke,
here the TLS Hamiltonian also remains unchanged.

\emph{(iv) Compression stroke.} This stage is accomplished by reversing
the protocol adopted in the above expansion stroke, such that the
TLS Hamiltonian is driven from $H_{h}=\frac{1}{2}\hbar\omega_{h}\sigma_{y}$
to $H_{c}=\frac{1}{2}\hbar\omega_{c}\sigma_{x}$ and the TLS state
evolves unitarily from to $\rho_{3}^{S}$ to $\rho_{4}=U^{\dagger}\rho_{3}^{S}U$.

The main quantity we are interested in is the QOHE efficiency $\eta_{TLS}$
. In the field of quantum thermodynamics, the engine efficiency is
given by $\eta=-\left\langle W_{ext}\right\rangle /\left\langle Q_{abs}\right\rangle $,
where $\left\langle Q_{abs}\right\rangle $ is the average total heat
absorbed and $\left\langle W_{ext}\right\rangle $ is the average
net work extracted from the engine. Note that $\left\langle Q_{abs}\right\rangle >0$
means energy flow into the engine, while $\left\langle W_{ext}\right\rangle <0$
means energy flow out of the engine. In order to determine the efficiency,
we need to introduce the first law of thermodynamics, together with
work and heat definitions. The first law of thermodynamics establishes
that the change in the internal energy of a given system during a
thermodynamic process can be decomposed into heat and work. In the
quantum thermodynamics domain, the first law is written as $\left\langle \Delta E\right\rangle =\left\langle Q\right\rangle +\left\langle W\right\rangle $,
where $\left\langle \Delta E\right\rangle $ is the average change
in internal energy, which is given by $\left\langle E\right\rangle =\text{Tr}\left(\rho H\right)$.
The definitions of heat and work averages we are interested in are
$\left\langle Q\right\rangle =\int dt\text{Tr}\left[\left(d\rho/dt\right)H\right]$
and $\left\langle W\right\rangle =\int dt\text{Tr}\left[\rho\left(dH/dt\right)\right]$
\cite{Alicki1979,Kosloff1984}, such that $\left\langle W\right\rangle =0$
($\left\langle \Delta E\right\rangle =\left\langle Q\right\rangle $)
in the heating and cooling strokes and $\left\langle Q\right\rangle =0$
($\left\langle \Delta E\right\rangle =\left\langle W\right\rangle $)
in the expansion and compression strokes.

We can now obtain the average heat exchanged with the reservoirs,
the average net work, and then the efficiency. Thus, with the information
provided in (i)-(iv) strokes, we obtain 
\begin{equation}
\left\langle Q_{h}^{S}\right\rangle =\frac{1}{2}\hbar\omega_{h}\left(\tanh\theta_{c}-\zeta\tanh\theta_{h}\right)-\hbar\xi\omega_{h}\tanh\theta_{c},
\end{equation}
\begin{equation}
\left\langle Q_{c}\right\rangle =-\frac{1}{2}\hbar\omega_{c}\left(\tanh\theta_{c}-\zeta\tanh\theta_{h}\right)-\hbar\xi\zeta\omega_{c}\tanh\theta_{h}
\end{equation}
and
\begin{multline}
\left\langle W_{net}\right\rangle =-\frac{1}{2}\hbar\left(\omega_{h}-\omega_{c}\right)\left(\tanh\theta_{c}-\zeta\tanh\theta_{h}\right)\\
+\hbar\xi\left(\omega_{h}\tanh\theta_{h}+\zeta\omega_{c}\tanh\theta_{h}\right),\label{Net Work}
\end{multline}
where $\theta_{c\left(h\right)}=\frac{1}{2}\beta_{c\left(h\right)}\hbar\omega_{c\left(h\right)}$,
$\zeta=1/\left(\mu{}^{2}+\nu{}^{2}\right)^{2}$ and $\xi=\left|\langle\pm_{y}\vert U\vert\mp_{x}\rangle\right|^{2}=\left|\langle\pm_{x}\vert U^{\dagger}\vert\mp_{y}\rangle\right|^{2}$.
Note that $\left|\pm_{x}\right\rangle $ is the eigenstate of the
Hamiltonian $H_{c}$ with eigenenergy $\pm\frac{1}{2}\hbar\omega_{c}$.
The transition probability $\xi$, which we call \emph{adiabaticity
parameter}, characterizes the time regime in which the engine operates:
$\xi=0$ corresponds to the quasi-static regime, while $\xi>0$ corresponds
to the finite-time regime. Imposing the work extraction condition
on the average net work, $\left\langle W_{net}\right\rangle <0$,
we achieve the condition

\begin{equation}
\xi<\frac{\left(\omega_{h}-\omega_{c}\right)\left(\tanh\theta_{c}-\zeta\tanh\theta_{h}\right)}{2\left(\omega_{h}\tanh\theta_{c}+\zeta\omega_{c}\tanh\theta_{h}\right)},\label{Ad. Parameter}
\end{equation}
which implies $\zeta<\tanh\theta_{c}/\tanh\theta_{h}$ (since $\xi\geq0$).
This condition results in heat absorption from the squeezed hot thermal
reservoir, $\left\langle Q_{h}^{S}\right\rangle >0$, and heat loss
to the cold thermal reservoir, $\left\langle Q_{c}\right\rangle <0$.
Therefore, the work extraction condition implies that $\left\langle W_{ext}\right\rangle =\left\langle W_{net}\right\rangle $
and $\left\langle Q_{abs}\right\rangle =\left\langle Q_{h}^{S}\right\rangle $,
which results in efficiency
\begin{equation}
\eta_{TLS}=1-\frac{\omega_{c}}{\omega_{h}}\left(\frac{1+2\xi{\cal F}}{1-2\xi{\cal G}}\right),\label{TLS Efficiency}
\end{equation}
 where ${\cal F}=\zeta\tanh\theta_{h}/\left(\tanh\theta_{c}-\zeta\tanh\theta_{h}\right)$
and ${\cal G}=\tanh\theta_{c}/\left(\tanh\theta_{c}-\zeta\tanh\theta_{h}\right)$.

In order to optimize the engine, we can maximize the average work
extracted by imposing the maximum power condition $d\left\langle W_{ext}\right\rangle /d\Delta\omega=0$,
where $\Delta\omega=\omega_{h}-\omega_{c}$. When applying the maximum
power condition to Eq. \eqref{Net Work} and considering $\theta_{c\left(h\right)}\ll1$,
which means that $\tanh\theta_{c\left(h\right)}\approx\theta_{c\left(h\right)}$,
we obtain the optimization ratio \textbf{
\begin{equation}
\frac{\omega_{c}}{\omega_{h}}=\frac{2\zeta\left(\frac{\beta_{h}}{\beta_{c}}\right)}{\left(1-2\xi\right)\left[1+\zeta\left(\frac{\beta_{h}}{\beta_{c}}\right)\right]}.\label{Opt. Condition}
\end{equation}
}Here we consider $\theta_{c\left(h\right)}\ll1$ due to the simplicity
it brings. Then, by replacing Eq. \eqref{Opt. Condition} in Eq. \eqref{TLS Efficiency},
we have the optimized efficiency\textbf{
\begin{equation}
\eta_{LTS}^{opt}=1-2\zeta\left(\frac{\beta_{h}}{\beta_{c}}\right)\left\{ \frac{2-\left(1-2\xi\right)^{2}\left[1+\zeta\left(\frac{\beta_{h}}{\beta_{c}}\right)\right]}{\left(1-2\xi\right)^{2}\left[1-\zeta^{2}\left(\frac{\beta_{h}}{\beta_{c}}\right)^{2}\right]}\right\} .\label{Opt. TLS Efficiency}
\end{equation}
}It is important to remember that the parameter $\xi$ in Eq. \eqref{Opt. TLS Efficiency}
allows us to study the efficiency $\eta_{LTS}^{opt}$ in any time
regime.

\begin{figure}
\begin{center}
\includegraphics{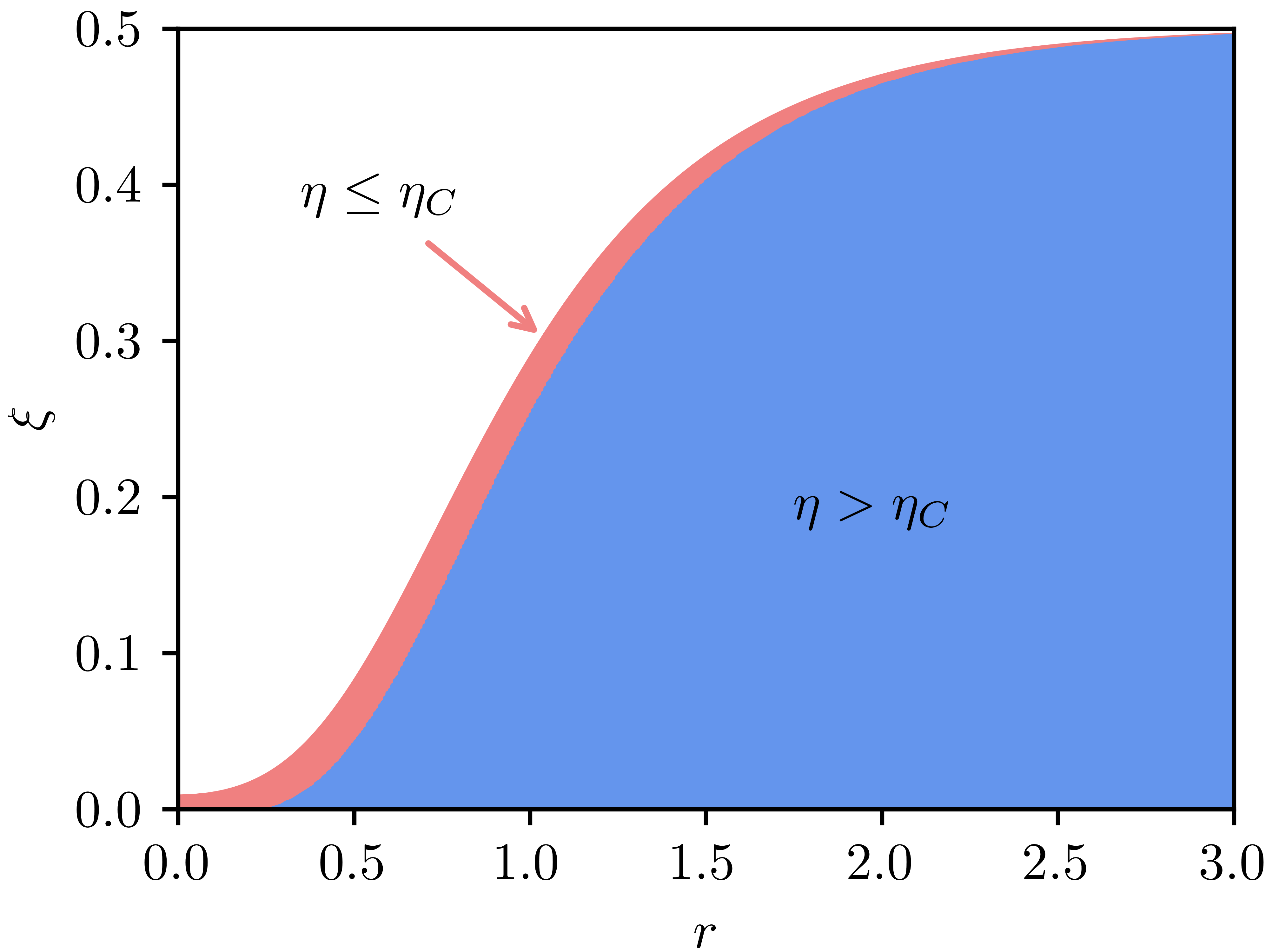}
\caption{\label{Fig1} Transition probability $\xi$ versus the squeeze parameter
$r$ . Blue and red regions separate the regimes where efficiency
is improved (blue region) as compared with the conventional Carnot
efficiency obtained using the QOHE model (red region).}
\end{center}
\end{figure}

\begin{figure}[t]
\begin{center}
\centering{}\includegraphics{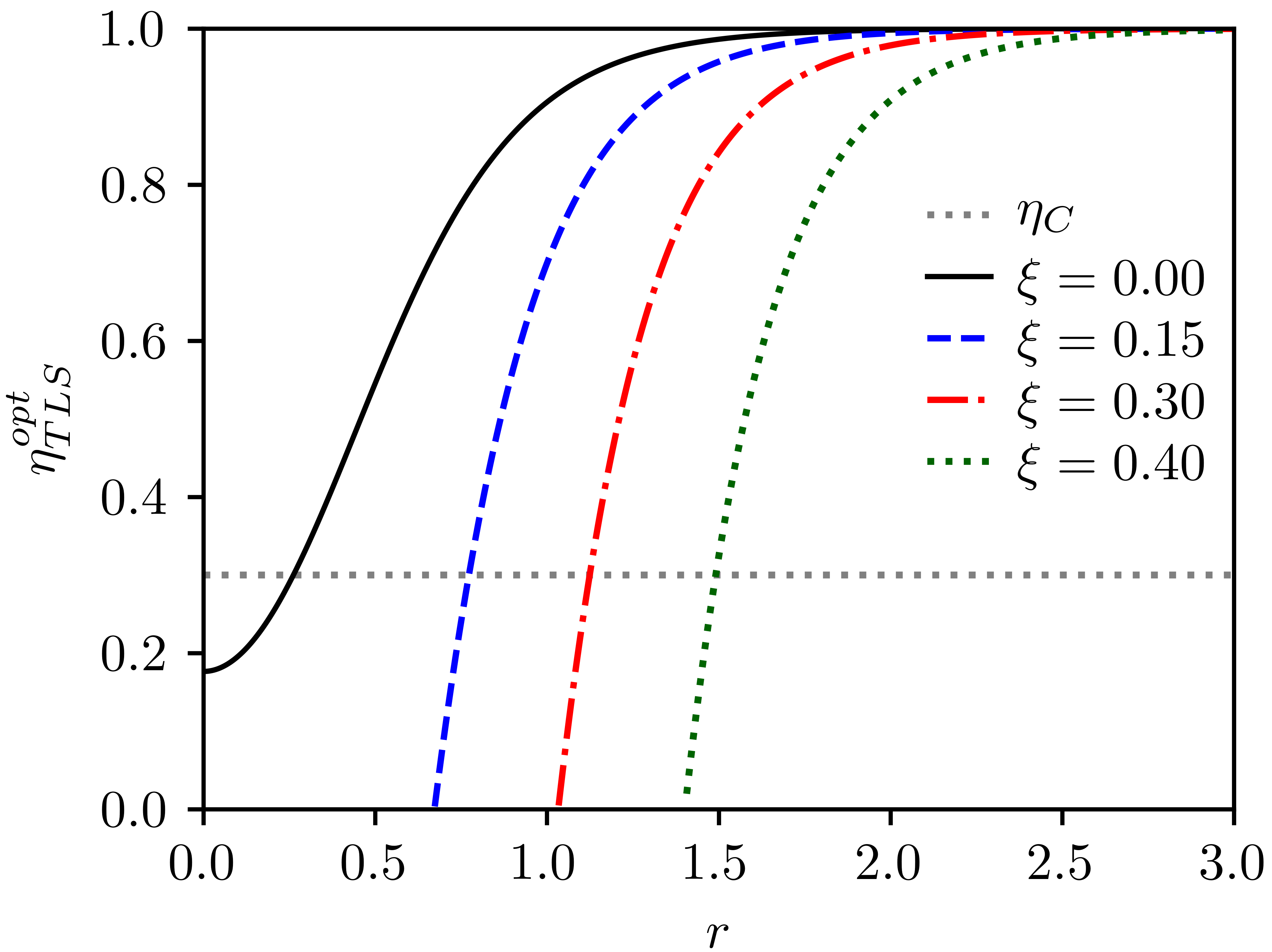}
\caption{\label{Fig2} Efficiency $\eta_{TLS}^{opt}$ as a function of the squeezing
parameter $r$ for $\beta_{h}/\beta_{c}=0.7$ and several transition
probabilities $\xi$. The solid black line corresponds to $\xi=0$,
the dashed blue line to $\xi=0.15$, the dash-dotted red line to $\xi=0.3$,
and the dotted green line to $\xi=0.4$. The horizontal dotted gray
line indicates the Carnot efficiency $\eta_{C}=1-\beta_{h}/\beta_{c}$.}
\end{center}
\end{figure}

So far, non-null power scenario was not fully explored when using
squeezed reservoirs. Thus, for the first time, we show how Carnot
efficiency can be surpassed even far from the quasi-static regime
with our TLS model. For this purpose, consider a specific case of
our engine, the case where $\beta_{h}/\beta_{c}=0.7$, to visualize
the behavior of $\eta_{LTS}^{opt}$ (Eq. \eqref{TLS Efficiency})
by varying both the squeezing $r$ and adiabaticity $\xi$ parameters.
For this case, Fig. \ref{Fig1} shows the values of $r$ and $\xi$
in which the net work is extracted from the engine: in the white region
there is no extraction of work, since Eq. \eqref{Ad. Parameter} is
not satisfied. The blue region indicates enhancement in the QOHE as
compared to the conventional Carnot engine, which is maximum at the
quasi-static regime. The narrow red region indicates where the Carnot
efficiency is not surpassed. Note that, even in the finite-time regime
($\xi>0$) the efficiency can surpass the Carnot efficiency. Besides,
in Fig. \ref{Fig2} we show the efficiency $\eta_{TLS}^{opt}$ \emph{versus}
the squeezing parameter $r$ for several values of the adiabaticity
parameter $\xi$. The optimized efficiencies are shown for $\xi=0$
(solid black line), $\xi=0.15$ (dashed blue line), $\xi=0.3$ (dash-dotted
red line), and $\xi=0.4$ (dotted green line). The Carnot efficiency
$\eta_{C}$ is indicated by the horizontal dotted gray line. We can
see in Fig. \ref{Fig2} that $\eta_{LTS}^{opt}$ can overcome $\eta_{C}$
and tends the unity to several different values of $\xi$. The important
point to be noted here is that we can now execute the expansion and
compression strokes quickly and still obtain high efficiencies, thus
bypassing the inconvenient internal friction (dissipation of useful
energy) that occurs with the decrease in $\tau$ \cite{Plastina2014,Cakmak2016,Peterson2019}.

\begin{figure}[t]
\begin{center}
\includegraphics{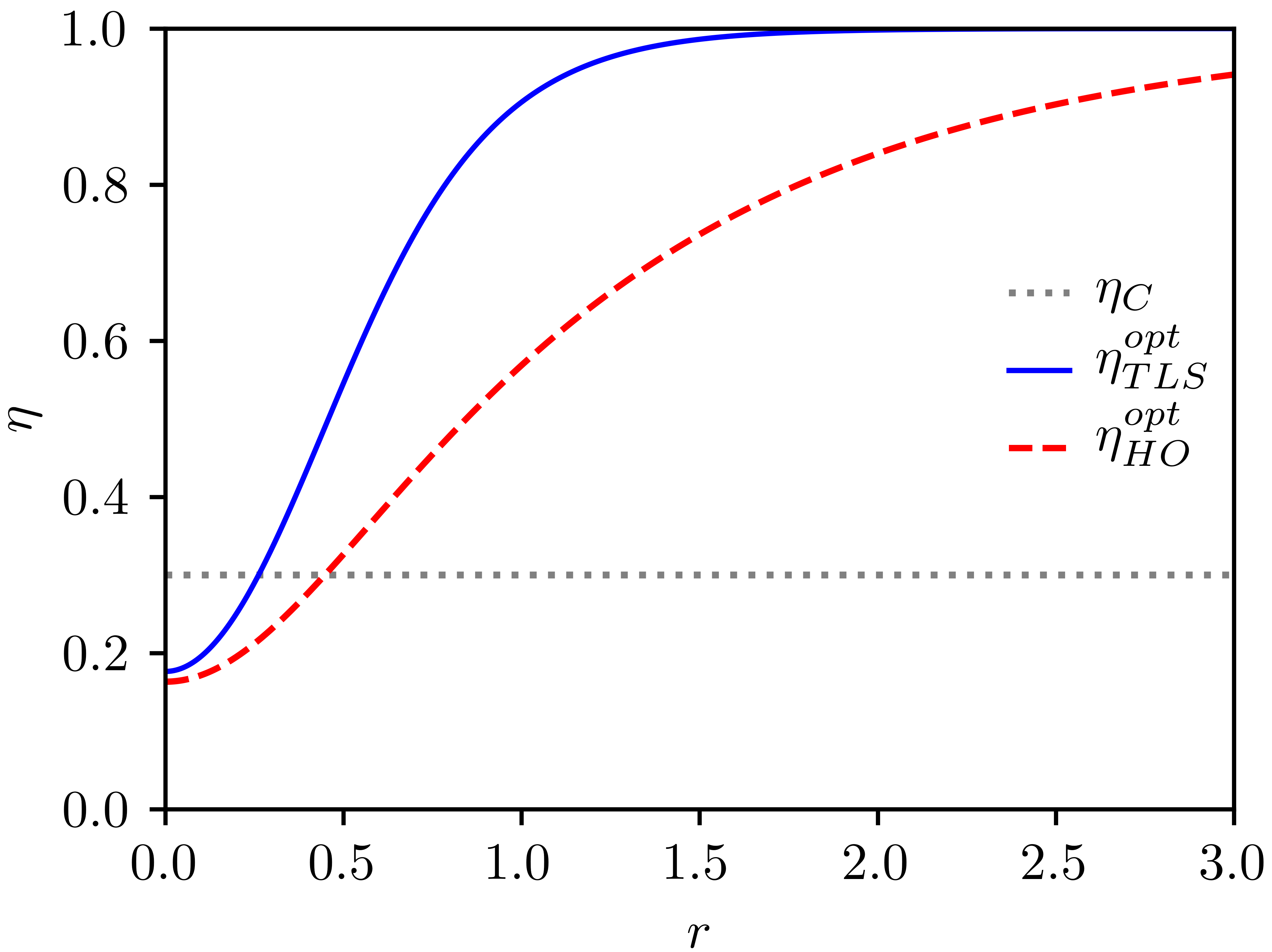}
\caption{\label{Fig3} Efficiencies $\eta_{TLS}^{opt}$ and $\eta_{HO}^{opt}$
as a function of the squeezing parameter $r$ considering quasi-static
processes, $\xi=0$. The temperature ratio is $\beta_{2}/\beta_{1}=0.7$.
The solid blue line is for the TLS model and the dashed red line is
for the HO model. The horizontal dotted gray line indicates the Carnot
efficiency $\eta_{C}=1-\beta_{h}/\beta_{c}$.}
\end{center}
\end{figure}

Now we compare our result with that obtained in Ref.~\onlinecite{RoBnagel2014}.
In that work, the authors address only the quasi-static regime of
a QOHE based on a HO\textbf{ }operating with a squeezed thermal reservoir.
The optimized efficiency obtained by them (Eq. (6) of Ref.~\onlinecite{RoBnagel2014},
takes the form

\begin{equation}
\eta_{HO}^{opt}=1-\sqrt{\zeta\left(\frac{\beta_{h}}{\beta_{c}}\right)}.\label{Opt. HO Efficiency (Quasi-static)}
\end{equation}
To our TLS model, replacing $\xi=0$ in Eq. \eqref{Opt. TLS Efficiency},
we have
\begin{equation}
\eta_{TLS}^{opt}=1-\frac{2\zeta\left(\frac{\beta_{h}}{\beta_{c}}\right)}{1+\zeta\left(\frac{\beta_{h}}{\beta_{c}}\right)}.\label{Opt. TLS Efficiency (Quasi-static)}
\end{equation}
To better appreciate the differences between the TLS and that of the
HO models, in Fig. \ref{Fig3} we show both the optimized efficiencies
$\eta_{TLS}^{opt}$ (solid blue line) and $\eta_{HO}^{opt}$ (dashed
red line) \emph{versus} the squeezing parameter \emph{$r$ }for the
ratio $\beta_{h}/\beta_{c}=0.7$. We can see that, for the same value
of the squeezing parameter, our model provides a larger efficiency.
This is a significant enhancement in the efficiency as compared with
Ref.~\onlinecite{RoBnagel2014}.

\begin{figure}[t]
\begin{center}
\includegraphics{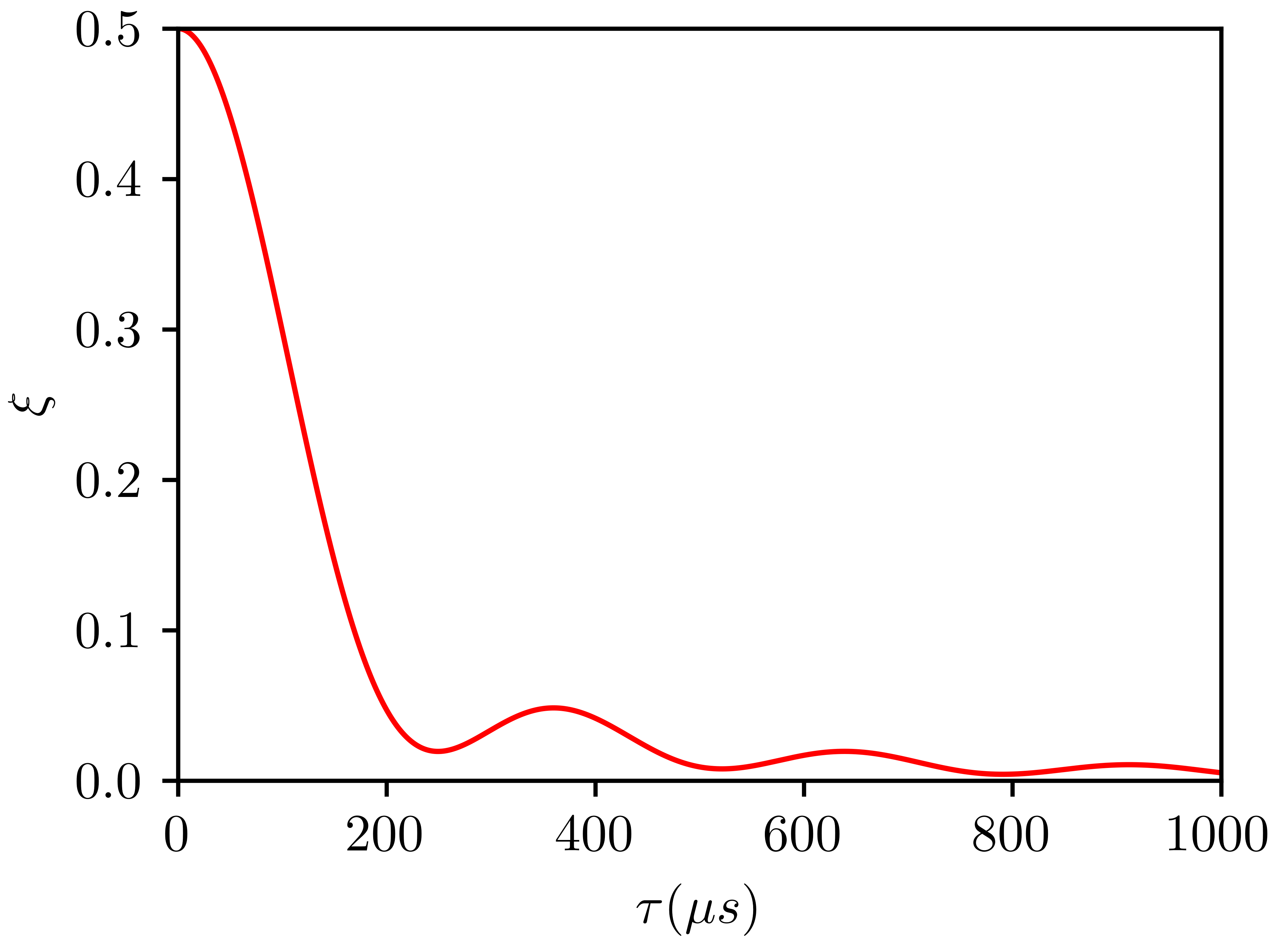}
\caption{\label{Fig4}The adiabaticity parameter $\xi$ as a function of the time
$\tau$ for the time-evolution operator $U={\cal T}\text{e}^{-\left(i/\hbar\right)\int_{0}^{\tau}H\left(t\right)dt}$,
where ${\cal T}$ is the time ordering operator and $H\left(t\right)=\frac{1}{2}\hbar\left[\omega_{c}\left(1-\frac{t}{\tau}\right)+\omega_{h}\frac{t}{\tau}\right]\left[\cos\left(\frac{\pi t}{2\tau}\right)\sigma_{x}+\sin\left(\frac{\pi t}{2\tau}\right)\sigma_{y}\right]$.
Here we used the following parameters: $\omega_{c}=2\pi\times2.5$
kHz, $\omega_{h}=10\omega_{c}$, $\beta_{c}=1/\left(10\ \text{peV}\right)$
, and $\beta_{h}=0.7\beta_{c}$. Note that the transition probability
vanishes for process occurring at null power.}
\end{center}
\end{figure}

\begin{figure}[t]
\begin{center}
\includegraphics{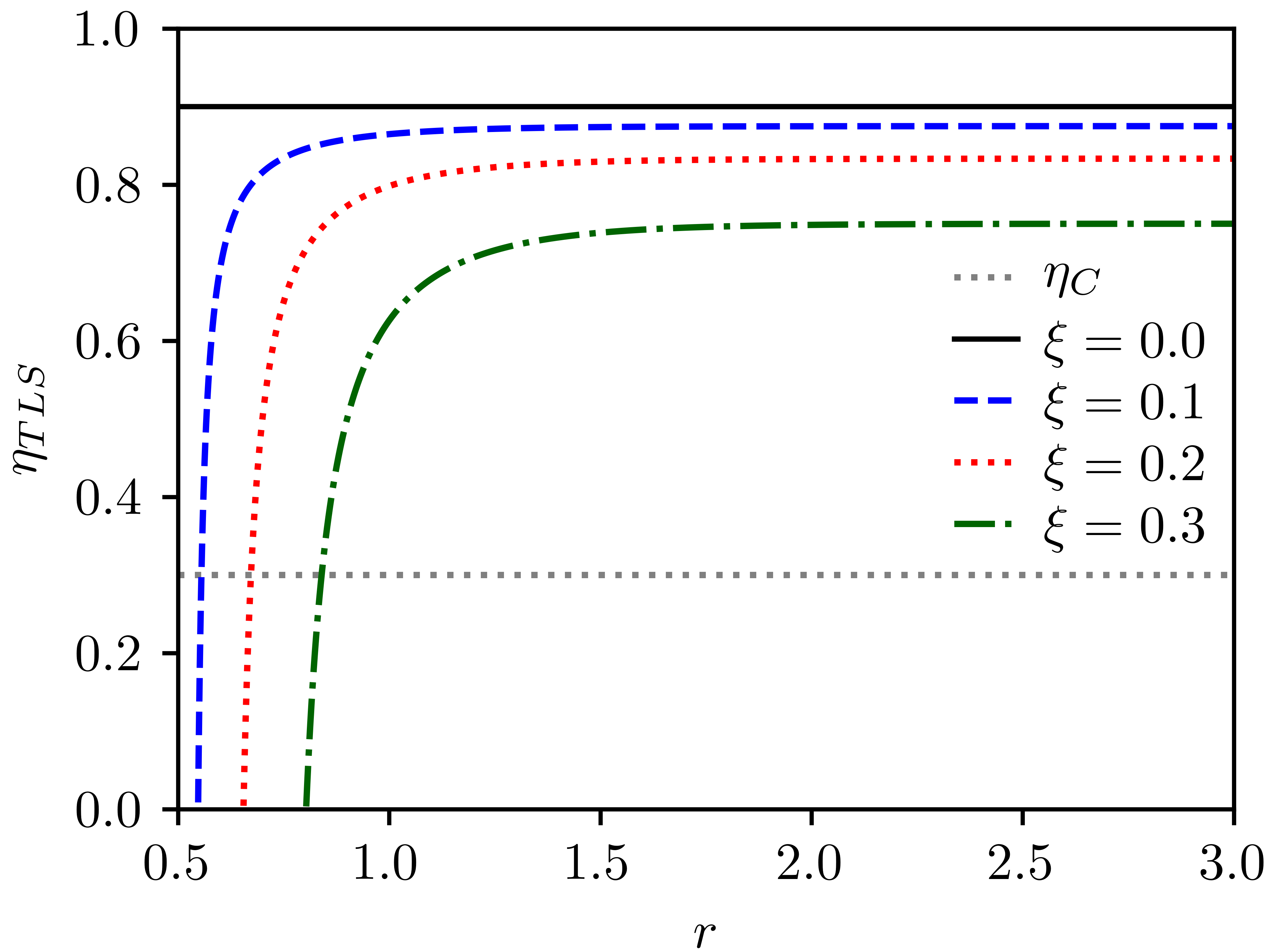}
\caption{\label{Fig5}Efficiency $\eta_{TLS}$ as a function of the squeezing
parameter $r$ considering quasi-static unitary processes, $\xi=0$
(solid black line), and non quasi-static unitary processes, $\xi=0.1$
(dashed blue line), $\xi=0.2$ (dotted red line) and $\xi=0.3$ (dash-dotted
green line). The parameters used are $\omega_{c}=2\pi\times2.5$ kHz,
$\omega_{h}=10\omega_{c}$, $\beta_{c}=1/\left(10\ \text{peV}\right)$
, and $\beta_{h}=0.7\beta_{c}$. The horizontal dotted gray line indicates
the Carnot efficiency $\eta_{C}=1-\beta_{h}/\beta_{c}$, which is
surpassed even for non quasi-static processes ($\xi>0$).}
\end{center}
\end{figure}

Next, aiming implementation in NMR \cite{Peterson2019,Assis2019},
we adopt the following parameters: $\omega_{c}=2\pi\times2.5$ kHz,
$\omega_{h}=10\omega_{c}$, $\beta_{c}=1/\left(10\ \text{peV}\right)$
, and $\beta_{h}=0.7\beta_{c}$. Since these parameters result in
$\theta_{c}\approx0.52$ and $\theta_{h}\approx3.62$, which means
that condition $\theta_{c\left(h\right)}\ll1$ is not satisfied, it
will be necessary to abandon Eq. \eqref{Opt. TLS Efficiency} and
extract numerical information directly from Eq. \eqref{TLS Efficiency}.
In the context of NMR, the time evolution operator to be considered
is $U={\cal T}\text{e}^{-\left(i/\hbar\right)\int_{0}^{\tau}H\left(t\right)dt}$,
with ${\cal T}$ being the time ordering operator and $H\left(t\right)=\frac{1}{2}\hbar\left[\omega_{c}\left(1-\frac{t}{\tau}\right)+\omega_{h}\frac{t}{\tau}\right]\left[\cos\left(\frac{\pi t}{2\tau}\right)\sigma_{x}+\sin\left(\frac{\pi t}{2\tau}\right)\sigma_{y}\right]$.
Fig. \ref{Fig4} shows the graph of the adiabaticity parameter $\xi$
as a function of time $\tau$ for the given time evolution operator.
In Fig. \ref{Fig5} we show the efficiency $\eta_{TLS}$ as a function
of the squeezing parameter $r$ for the adiabatic parameters $\xi=0$
(solid black line), $\xi=0.1$ (dashed blue line), $\xi=0.2$ (dotted
red line), and $\xi=0.3$ (dash-dotted green line). For comparison,
Fig. \ref{Fig5} also shows the Carnot efficiency $\eta_{C}=1-\beta_{h}/\beta_{c}$
(dotted gray line). As we can see, the experimental parameters available,
although impose limit to the efficiency attained, allow to obtain
efficiencies surpassing Carnot efficiency even for processes occurring
far from quasi-static regime.

\section*{Conclusion}

We have proposed a quantum Otto heat engine based on a two-level system
as the working substance that operates under two reservoirs: a cold
thermal reservoir and a squeezed hot thermal reservoir. While for
typical scenarios Carnot efficiency is the higher limit and can be
achieved only for quasi-static processes, in this work we have demonstrated
that it is possible to surpass Carnot efficiency even in the finite-time
regime (at non null power). Also, we showed that
given the same squeezing parameter and the same temperature ratios,
the two-level system model exhibits higher efficiency at maximum power
as compared to the harmonic oscillator model.

\section*{Acknowledgments }

We acknowledge financial support from the Brazilian agencies, CAPES
(Financial code 001) CNPq and FAPEG. This work was performed as part
of the Brazilian National Institute of Science and Technology (INCT)
for Quantum Information Grant No. 465469/2014-0.


%

\end{document}